# The exchange-correlation potential correction to the vacuum potential barrier of graphene edge

Weiliang Wang, Junwen Shao and Zhibing Li[1]

State Key Laboratory of Optoelectronic Materials and Technologies, School of Physics and Engineering, Sun Yat-sen University, Guangzhou, P.R. China 510275

Email: stslzb@mail.sysu.edu.cn

**Abstract.** We investigated the vacuum potential barriers of various graphene edges terminated with hydrogen, oxygen, hydroxyl group and ether group respectively. It is found that the exchange-correlation potential correction is significant to the edge structures with electronegativity higher than carbon. The correction leads to the local work function decreased by more than 1. eV for the O terminated edge, while the local work function of H and OH terminated edges remain unchanged.

Keyword: exchange-correlation potential, local work function, graphene

PACS: 71.15.Mb, 79.70.+q, 73.30.+y

## 1. Introduction

The vacuum potential barrier is a surface characteristic of materials that is crucial for photoemission, field emission, and thermo-emission. The field emission is particularly sensitive to the detail of the vacuum potential barrier of the surface/edge of the emitter because field emission is an electron tunneling assisted via electric field. The simplest model for the barrier of a bulk metal (semiconductor) is a step function with vacuum potential energy higher than the Fermi level (bottom of conduction band) by a constant called local work function (local electron affinity). More exactly, the potential energy is not a simple function of spatial coordinates in the vicinity of the surface and an exchange-correlation potential (XCP) should be included even outside the surface of atomic lattice [1], but it is common practice to approximate this by a classical image potential. The total electric potential energy is equal to the work done by the electric field in a process of moving the testing electron from the concerning position to the anode. In macroscopic scale this potential is easy to calculate with electrostatics [2]. For nanostructures, generally one can not define a unique local work function (or electron affinity) because the scale of surface/edge may be comparable with the atomic scale and the electric field caused by the dipole moment (hereinafter referred to as polarization) at the surface/edge generally is not perpendicular to the surface/edge, therefore the potential would be varying significantly along the surface/edge [3]. A question is how much the surface/edge charge polarization will change the local vacuum potential barrier? It should be answered with more sophisticated method such as the density functional theory (DFT). One may first use DFT to calculate the charge density and then integrate the Coulomb potential of each source charge to give the total electric potential. This method yields a better approximation for the vacuum potential barrier [4-16]. However this approximation has a paradox: if the surface/edge is negatively polarized, i.e., with the negative edges of surface/edge dipoles outwards, it would imply the electrons have more probability to go beyond the atomic lattice, but on the other hand more negative charges at the surface/edge would raise the vacuum potential barrier more thus would decrease the probability of electrons to leave the atomic lattice.

---

[1] Corresponding author at: School of Physics and Engineering, Sun Yat-sen University, Guangzhou, P.R. China 510275. Email: stslzb@mail.sysu.edu.cn Tel:+86-2084111107, Fax: +86-2084112878

We solve this paradox by noting that the electric potential energy is not the total potential barrier for emitting electrons. There is the exchange-correlation potential (XCP) in the region where the electron wave function of the ground state is not negligible. The XCP is a many electron effect that is not included in the electric potential. There have been many discussions on and models for the XC in the frame of DFT. It is known that the XCP being a functional of electron density is crucial for a successful prediction of atomic and electronic structures of the material [17]. The XCP effects on the surface barrier have been much discussed in the context of metal surfaces [1] . But in the context of nanostructures, such as carbon nanotubes and graphene, there is rarely any discussion on the effect of XCP on the vacuum potential barrier. We investigate graphenes with various edge structures as examples to show correction of XCP and its dependence on the edge structures. The correction of XCP turns out to be large for some edge structures and negligible for some others.

**2. Calculation method**
The total potential barrier is defined as the effective potential for an electron, which is the sum of the electric potential and the XCP. We first adopt the DFT (implemented in VASP) to calculate electron density of a graphene ribbon assuming the ribbon width is large enough (12~14 columns of carbons) such that the edge-edge interaction is negligible. Then the electric potential is calculated as in [18]. According to the local density approximation (LDA) [19], the XCP is uniquely determined via the density of electron. The electron density in the barrier region may be written as

$$n = A\exp(-x/\lambda) \qquad (1)$$

where $x$ is position along the path normal to the edge and parallel to the graphene plane with the origin at the outer most O or C atom of the edge. The factor A and the characteristic length $\lambda$ are two parameters determined by the DFT calculations, as presented in Table 1 for various edges that are described in the followings.

Table 1 Parameters in Eq. (1), electron density at the inner turning point $n_{turn}$, and local work function with and without XCP. (Local work function is defined as the difference between the

maximum of the potential barrier and the Fermi level.)

|  | A ($\text{Å}^{-3}$) | $\lambda$ (nm) | $n_{turn}$ ($\text{Å}^{-3}$) | local work function without XCP (eV) | local work function with XCP (eV) |
|---|---|---|---|---|---|
| Reconstructed Z-edge | 1.74 | 0.037 | 0.0051 | 5.43 | 5.04 |
| Z-edge + H | 1.64 | 0.043 | 0.0015 | 4.29 | 4.29 |
| Z-edge + OH | 7.82 | 0.027 | 0.0009 | 3.76 | 3.76 |
| Z-edge + O | 7.77 | 0.024 | 0.0120 | 7.74 | 6.62 |
| Z-edge + ether group | 7.74 | 0.025 | 0.0081 | 4.32 | 3.96 |
| Clean A-edge | 1.68 | 0.037 | 0.0033 | 4.91 | 4.71 |
| A-edge + H | 1.59 | 0.038 | 0.0014 | 4.32 | 4.32 |
| A-edge + O | 7.96 | 0.025 | 0.0095 | 7.11 | 5.93 |
| A-edge + 2O | 7.83 | 0.026 | 0.0133 | 7.26 | 6.19 |
| A-edge + ether group | 7.80 | 0.025 | 0.0050 | 5.97 | 5.34 |

The edge structures to be considered are presented in figure1. Figure 1 (a) to (e) are respectively the reconstructed zigzag edge (Z-edge), Z-edges with dangling bonds saturated by hydrogen, hydroxyl group, oxygen and ether group. Figure 1(f) to (j) are respectively the clean armchair edge (A-edge), the A-edge with dangling bonds saturated by hydrogen, with half/all the dangling bonds terminated by oxygen and ether group. The Allen electronegativities of hydrogen, carbon, and oxygen are 2.300, 2.544, and 3.610 respectively. In comparison with carbon, hydrogen has lower electronegativity, oxygen has higher electronegativity. The difference of these two classes is that the former has fewer electrons along the edge than the latter. Thus the dipoles from the edge polarization are directing outward/inward in the less/larger electronegativity class. Our calculations show that the reconstructed/clean edges have more electrons in the last column of carbons than other carbon atoms, so can be included in the class of higher electronegativity.

We adopted the Perdew-Zunger XCP [20]

$$\mu_{xc} = \begin{cases} -\frac{4}{3} 0.9164/r_s + \gamma \frac{1 + \frac{7}{6}\beta_1\sqrt{r_s} + \frac{4}{3}\beta_2 r_s}{\left(1 + \beta_1\sqrt{r_s} + \beta_2 r_s\right)^2} & r_s \geq 1 \\ -\frac{4}{3} 0.9164/r_s + A\ln r_s + B - \frac{A}{3} + \frac{2}{3}Cr_s \ln r_s - \frac{1}{3}(2D - C)r_s & r_s \leq 1 \end{cases} \quad (2)$$

Where $r_s = \left(\frac{3}{4\pi n}\right)^{1/3}$, $\gamma = -0.1423, \beta_1 = 1.0529, \beta_2 = 0.3334$, A = 0.0311, B = -0.048, C = 0.0020, D = -0.0116 (atomic unit). With electron density (1), one can find $r_s$ easily. Substituting it into (2) we obtain the XCP to add to the electric potential to form the vacuum potential barrier.

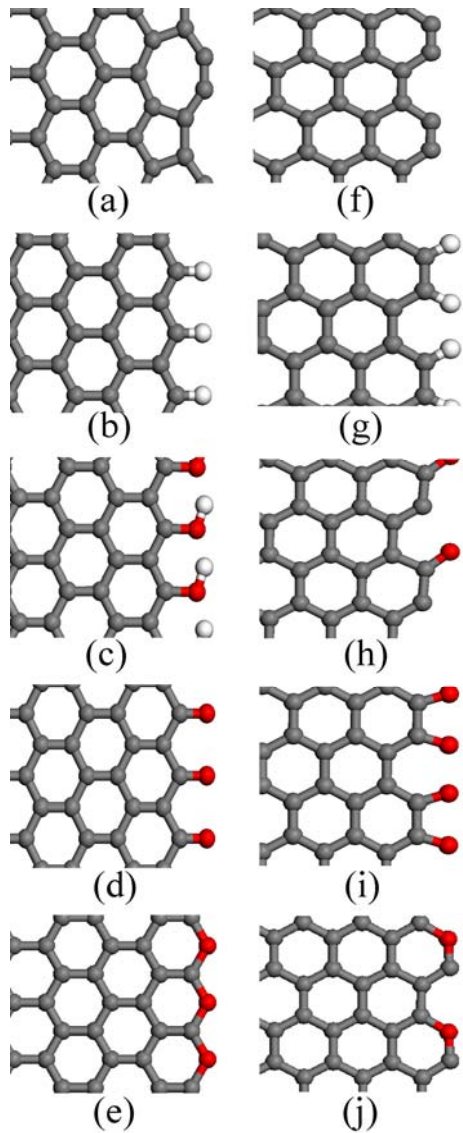

Figure 1. (Colour online) Edge structures: (a) reconstructed clean Z-edge [21]; (b) H terminated Z-edge; (c) OH terminated Z-edge; (d) O terminated Z-edge; (e) ether group terminated Z-edge; (f) clean A-edge; (g) H terminated A-edge; (h) half O terminated A-edge; (i) fully O terminated A-edge; (j) ether group terminated A-edge.

### 3. Results and discussion

The vacuum potential barriers, without (solid curves) and with (dashed curves) the XCP, for edge structures of figure 1(a)-(e) are compared in figure 2, while that for structures of figure 1(e)-(j) are in figure 3. First we note that the vacuum potential energy without the XCP has an obvious peak in the vicinity of an edge with larger electronegativity (figure 2 and figure 3). The reason is that electron has more probability to go beyond the atomic lattice in this class of edges and forms edge dipoles with outward negative ends that raise the electric potential energy in the vicinity of the edge but is ignorable in large distance away from the edge. The edges with less electronegativity have edge dipoles with out-ward positive ends; hence the barriers of them have no peak. The numerical results show that the peaks can be more than 1. eV (edges terminated by oxygen atoms). Therefore electron should be impossible to emit from these edges if the XCP could be ignored. This is the paradox we mentioned previously. Now we see that the XCP erases the unwanted peaks of electric potential barriers and the paradox is solved (dashed curves of figure 2 and figure 3). Another notable fact is that the XCP shifts the inner classical turning point outwards by a distance of order λ. This is reasonable because λ is just the characteristic length of the electron wave at the edge. In Table 1, the local work functions (defined as the difference between the Fermi level and the maximum of the potential barrier) with and without the XCP are compared. The XCP corrections to the local work functions are large for the edges of large electronegativity. To the edges of less electronegativity, the XCP corrections are negligible. The electron density at the inner classical turning point has been listed in Table 1. The electron densities of the lower electronegativity edges are about one order smaller than those of the higher electronegativity, confirming the correlation between the electron density and the XCP effect.

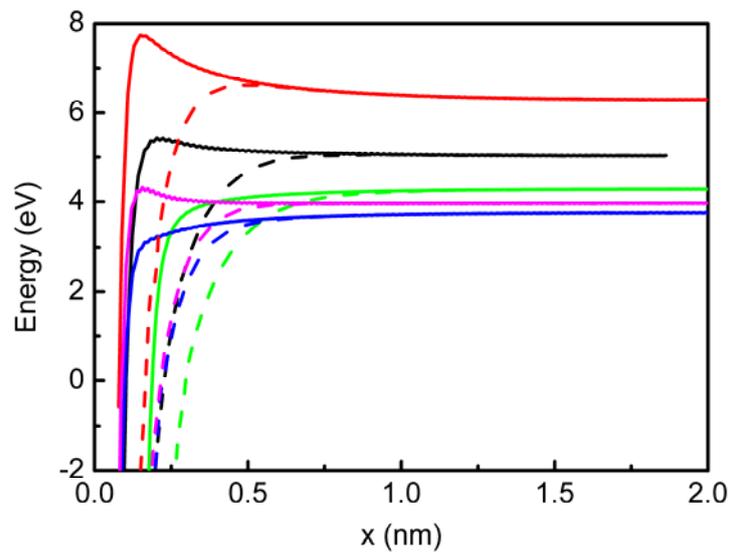

Figure 2. (Colour online) From top to bottom: Potential energies in vacuum in the vicinity of O terminated (red), reconstructed (black), H terminated (green), ether group terminated (magenta) and OH terminated (blue) Z-edge of graphene. The energies are shifted to let the Fermi level be zero. It is plotted along a line perpendicular to the edge and parallel to the graphene plane with the origin at the outer most O or C atom.

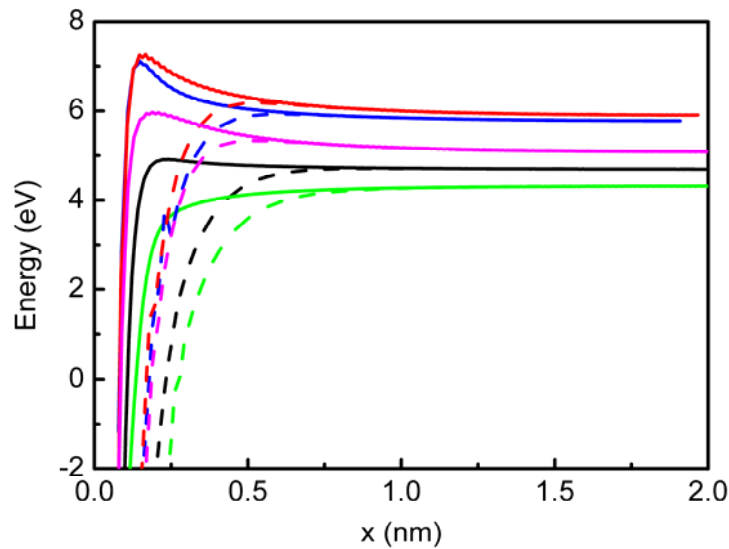

Figure 3. (Colour online) From top to bottom: Potential energies in vacuum in the vicinity of fully O terminated (red), half O terminated (blue), ether group terminated (magenta), clean (black) and H terminated (green) A-edge of graphene. The energies are shifted to let the Fermi level be zero. It is plotted along a line perpendicular to the edge and parallel to the graphene plane with the origin at the outer most O or C atom.

**4. Summary**

We calculated the exchange-correlation potential at various graphene edges. It suggests a general trend that edges terminated with atoms (or atom groups) of higher electronegativity will receive larger correction from the exchange-correlation potential. It is found that the exchange-correlation potential shifts the inner classical turning points of all edges to the vacuum side. Remarkably, the exchange-correlation potential erases the electric potential peaks of the edge structures that are more electronegative than carbon. We found that the zigzag edge terminated with the ether group has local work function 3.96eV that is lower than that of the hydrogen terminated zigzag edge (4.29eV) and armchair edge (4.32eV). The effects of the exchange-correlation potential on the vacuum potential barrier should appear generally in nanostructures where the surface/edge charge polarization is crucial. For instance, it seems demanding to reinvestigate the vacuum potential barrier of carbon nanotubes with the correction of the exchange-correlation potential [22-28].

**Acknowledgement:**

The project is supported by the National Basic Research Programme of China (2007CB935500), the National Natural Science Foundation of China (Grant No. 11104358) and the high-performance grid computing platform of Sun Yat-sen University.